\documentclass{ws-rv9x6}
\usepackage{subfigure}   
\usepackage{ws-rv-thm}   
\usepackage{ws-rv-van}   
\makeindex
\usepackage{xspace}

\newcommand{\pt}{\ensuremath{{p_{\rm T}}}}

\newcommand{\Pythia}{{\sc Pythia}\xspace}
\newcommand{\EPOS}{{\sc EPOS}\xspace}

\newcommand{\QGSJET}{{\sc QGSJET}\xspace}

\begin{document}

\chapter[Measurement of Minimum Bias Observables with ATLAS]{Measurement of Minimum Bias Observables with ATLAS}


\author[J. Kvita]{Jiri Kvita\footnote{On behalf of the ATLAS collaboration.}}
\address{
Joint Laboratory for Optics, Palacky University\\
17. listopadu 50A, 772 07 Olomouc, Czech Republic,\\ 
jiri.kvita@upol.cz\footnote{Palacky University in Olomouc, Czech Republic.}}

\begin{abstract}
  The modelling of minimum bias interactions is a crucial ingredient to learn about the description of soft QCD processes. It has also a significant relevance for the simulation of the environment at the LHC with many concurrent $pp$ interactions (“pileup”). The ATLAS collaboration has provided new measurements of the inclusive charged-particle multiplicity and its dependence on transverse momentum and pseudorapidity in special data sets with low LHC beam currents, recorded at center of mass energies of 8 TeV and 13 TeV.
  The measurements use charged-particle selections with minimum transverse momentum of both 100~MeV and 500~MeV, and  various phase-space regions characterized by low and high charged-particle multiplicities.
\end{abstract}

\section{Introduction}

Measurements of charged-particle multiplicities are an important input for pile-up modelling in $pp$ collisions and a handle on multi-parton interactions at the LHC.
Such measurements provide insight into many aspects of non-perturbative quantum chromodynamics (QCD) in hadron collisions.
Measurements of observables sensitive to the modelling of minimum bias (MB) processes at several centre-of-mass energies test models predictions up to $O(10)$~TeV scales, providing also a connection to cosmic ray physic.
\section{ATLAS Inner Detector}

The ATLAS experiment~\cite{1748-0221-3-08-S08003} at the LHC~\cite{1748-0221-3-08-S08001} is a multi-purpose detector recording events in proton-proton collisions. It consists of several sub-systems focusing on particles tracking, calorimetry and identification.
Most importantly for the presented results, charged particles are reconstructed by a~set of silicon pixel and strip, and transition radiation detectors~\cite{Aad:2010bx}, providing the transverse momentum measurement, excellent position and vertex resolution and particle identification capabilities, enabling the discrimination between primary and secondary particles created in interactions in material or via decays.
An additional layer of pixel detectors was inserted for 13~TeV collisions, called the Insertable $b$-layer (IBL)~\cite{IBL}, allowed by a novel reduced-radius beam pipe, which is important for $b$-tagging and vertex resolution, with a reduced pixel size of  $50 \times 250\,\mu$m.

\section{Sample and Event Selection}

Minimum-bias trigger scintillators (MBTS) detectors consist of two octants connected to photomultipliers, covering a rapidity\footnote{ATLAS uses a
right-handed coordinate system with its origin at the nominal
interaction point (IP) in the centre of the detector and the $z$-axis
along the beam pipe. The $x$-axis points from the IP to the centre of
the LHC ring, and the $y$-axis points upward. Cylindrical coordinates
($r$,$\phi$) are used in the transverse plane, $\phi$ being the azimuthal angle around the beam pipe. The pseudorapidity is defined in terms of the
polar angle~$\theta$ as $\eta = - \ln \tan(\theta/2)$.
}
range from 2.1 to 3.8, which are highly efficient and radiation-hard. 
Special LHC fills at low pile-up were used. 
The recorded data include contributions from single- and double-diffractive processes as well as non-diffractive, which populate different rapidity and particle multiplicity regions.

\section{Observables}

All analyses presented here at 8 and 13 TeV centre-of-mass energy of the LHC measure the following spectra:
charged-particles multiplicity ($n_{\rm ch}$) as function of rapidity ($\eta$) and transverse momentum ($\pt$) of the leading track, and the tracks average transverse momentum in bins of $n_{\rm ch}$.
All distributions are corrected (unfolded) to stable-particles level, where a stable particle means its lifetime $\tau> 300$~ps (i.e. $c \tau> 9$~cm), thus including decay products of particles with $\tau < 30$~ps (i.e. $c\tau< 9$~mm).
Newly, particles with $30 < \tau < 300$~ps are excluded as these are mostly charged strange baryons with a low (0.3\%) reconstruction efficiency due to their late decay and fewer hits in silicon detectors. Inconsistencies in their description among models also yields a larger systematics uncertainty.

\section{Selection}

\subsection{Vertex Selection}
The primary vertex (PV) is required to be formed by at least two tracks of $\pt{} > 100$~MeV.
Events are vetoed if another PV is reconstructed from four or more tracks.
Events with an additional PV reconstructed from three or two tracks are kept as such are often a split PV or a secondary vertex reconstructed as a~PV.
\subsection{Tracks Selection and Efficiency}
Tracks within $|\eta| < 2.5$ and the transverse impact parameter ($d_0$) fulfilling $|d_0| < 1.5$~mm are considered, with additional hit requirements in pixel and strip detectors. Further track quality requirements are imposed, including a requirement on a~good $\chi^2$ for high-$\pt$ tracks, and a hit in the first silicon layer if expected. Last, at least one track is required for the $\pt{} > 500$~MeV analysis while $\geq 2$ track are required for the $\pt{} > 100$~MeV analysis.

\section{Data Corrections}
Data events are weighted in order to compensate for the decrease in the event yield due to finite trigger and vertex efficiencies. Each selected track is weighted to compensate for finite tracking efficiency and the expected fraction of background from non-prompt particles and strange baryons (predicted by \EPOS{}); and for the finite efficiency of tracks to be reconstructed from out-of-kinematic region.
A~Bayesian unfolding \cite{DAgostini:1994fjx} is applied to correct the measured distributions at the detector level to those at the stale-particles level.
With respect to the~7 TeV results, the novel 8 TeV measurements rely on~a new tracking algorithm and  better material description, leading to smaller systematic uncertainties. In addition, more fiducial regions of particles multiplicities are presented, with more models-constraining results.

\section{Systematic uncertainties}
 
Systematic variations of different modelling assumptions affect the corrections applied to the data distributions.
Scaling by the trigger, tracking and vertexing efficiencies affects distributions before the unfolding, and the uncertainties on the corresponding scale factors need to be propagated. 
Different shapes of particles spectra in data and simulation lead to a different efficiency, being a 2\% effect on average and utmost 5\% in high multiplicities.
Material description in simulation amounts to about a 5\% uncertainty at 8 TeV.
Secondaries fraction scaling between data and MC is a minor source due to low background levels.
Particles composition affects efficiency dependence on the particle type and is a~1\% effect.
Other minor sources of uncertainties include the momentum resolution which is negligible except on \pt{} spectra, and unfolding non-closure (1\%).
\section{Results}

In this contribution we will describe the 8 and 13 TeV results, referring reader to reference \cite{Aad:2010ac} for 0.9, 2.36 and 7~TeV results.

\subsection{8 TeV Results}

For the 8 TeV results~\cite{Aad:2016xww} with the $\pt{} > 500\,$MeV track requirement in the $n_{\rm ch} \geq 1$ phase space, the data set of 9M events, corresponding to the integrated luminosity of $160\,\mu{\rm b}^{-1}$, uses track impact parameters reconstructed w.r.t. beam spot (BS). It is found that the per-event charge-particles multiplicity is best described by the \EPOS{} generator. All generators describe the $\eta$ shape well while the \QGSJET{} generator does not describe the $\pt{}$ dependence, and also deviates at large multiplicities. It is worth noticing that the \Pythia{}8 A2 tune is derived based on the 7~TeV data while \Pythia{}8 Monash is a more general-purpose tune including SPS and Tevatron data.
Additional phase spaces of different charged-particles multiplicities of
$n_{\rm ch} \geq 2$, $n_{\rm ch} \geq 6$, $n_{\rm ch} \geq 20$, $n_{\rm ch} \geq 50$ were also studied, with a very good agreement to the \Pythia{}8 Monash tune but also to \EPOS{} and less of the \Pythia{} A2 tune while the \QGSJET{} generator fails to describe the data. 

\begin{figure*}[htbp]
  \centering
  \subfigure[]{\includegraphics[width=0.47\textwidth]{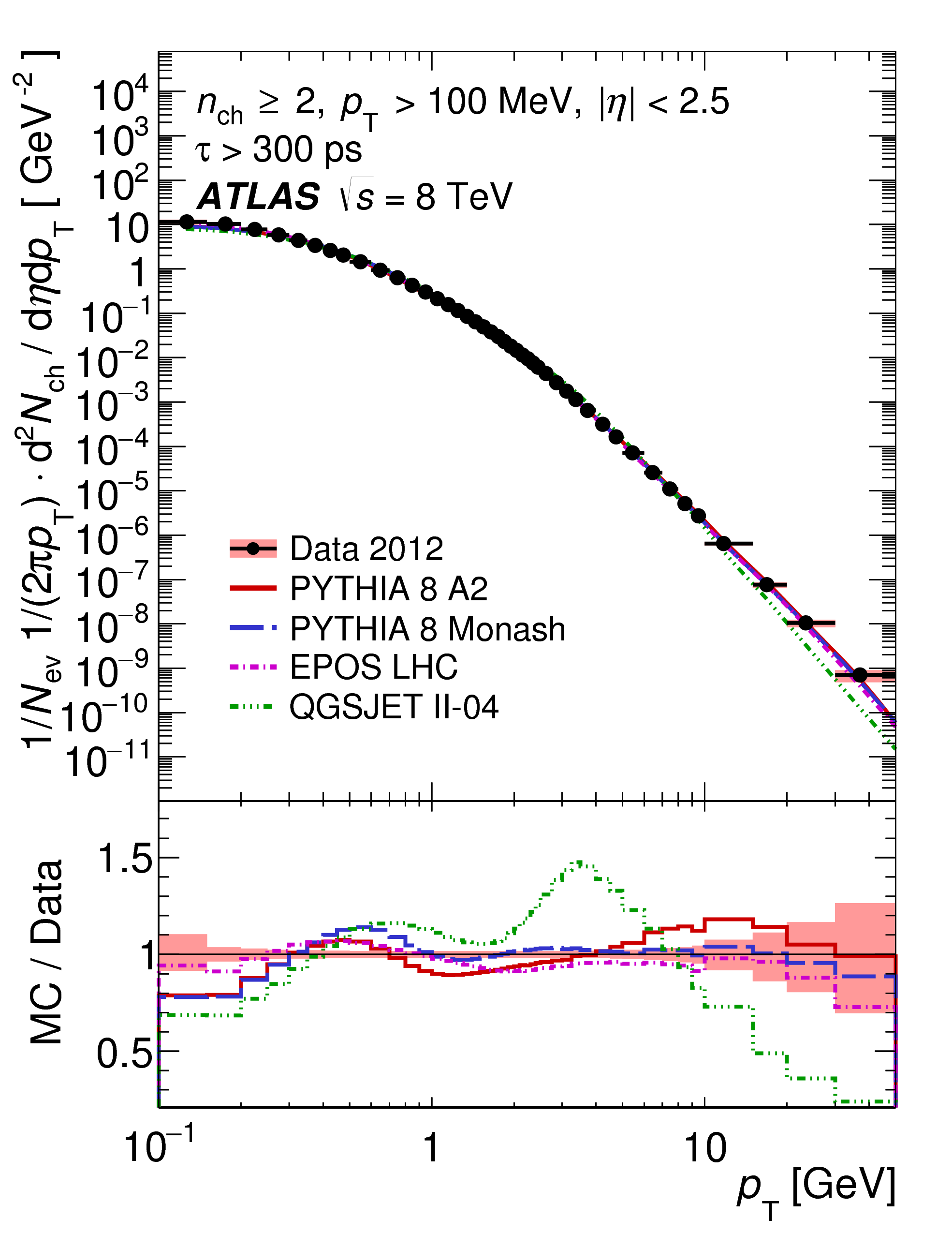}\label{fig:8sub4b}}
  \subfigure[]{\includegraphics[width=0.47\textwidth]{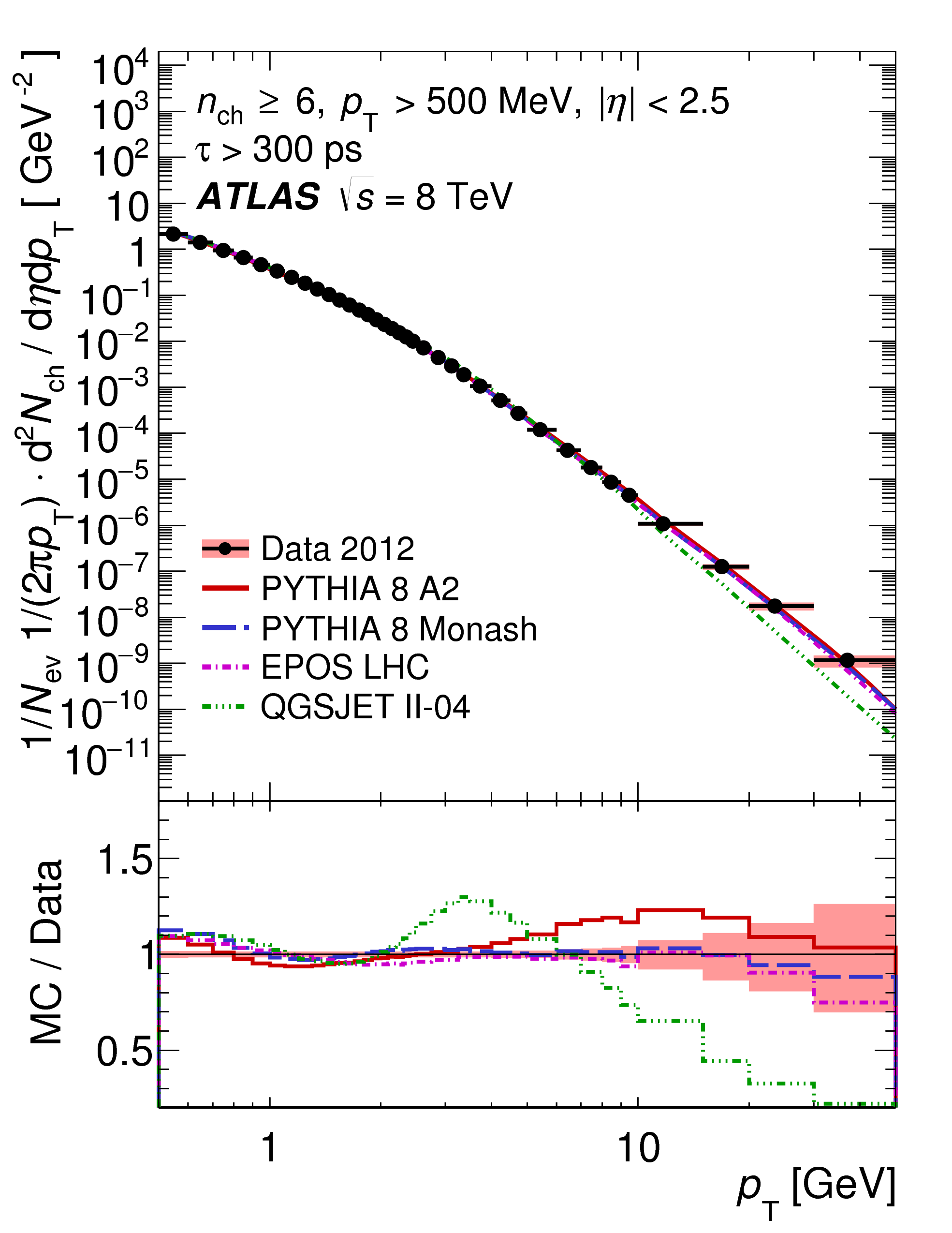}\label{fig:8sub6b}}
  \subfigure[]{\includegraphics[width=0.47\textwidth]{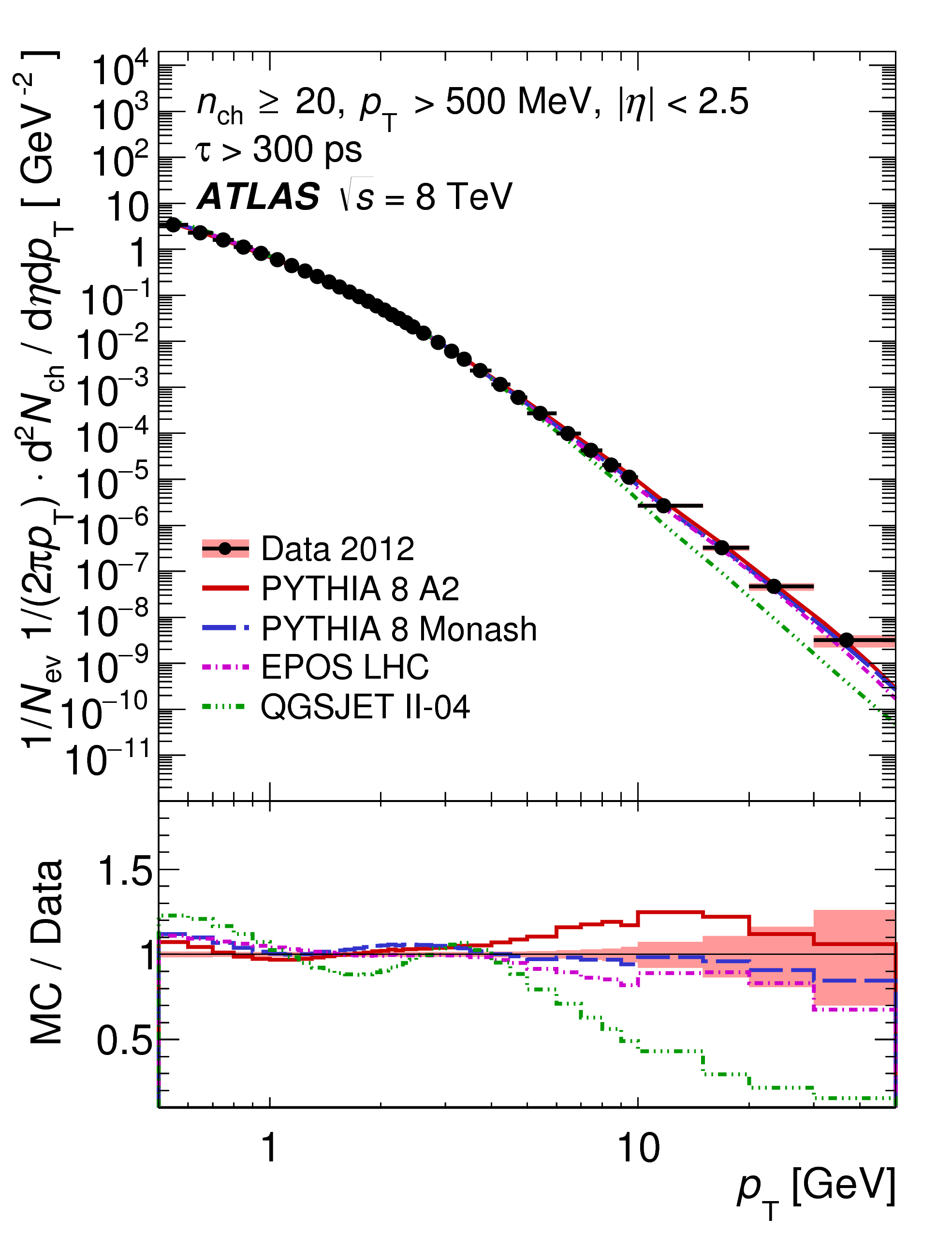}\label{fig:8sub7b}}
  \subfigure[]{\includegraphics[width=0.47\textwidth]{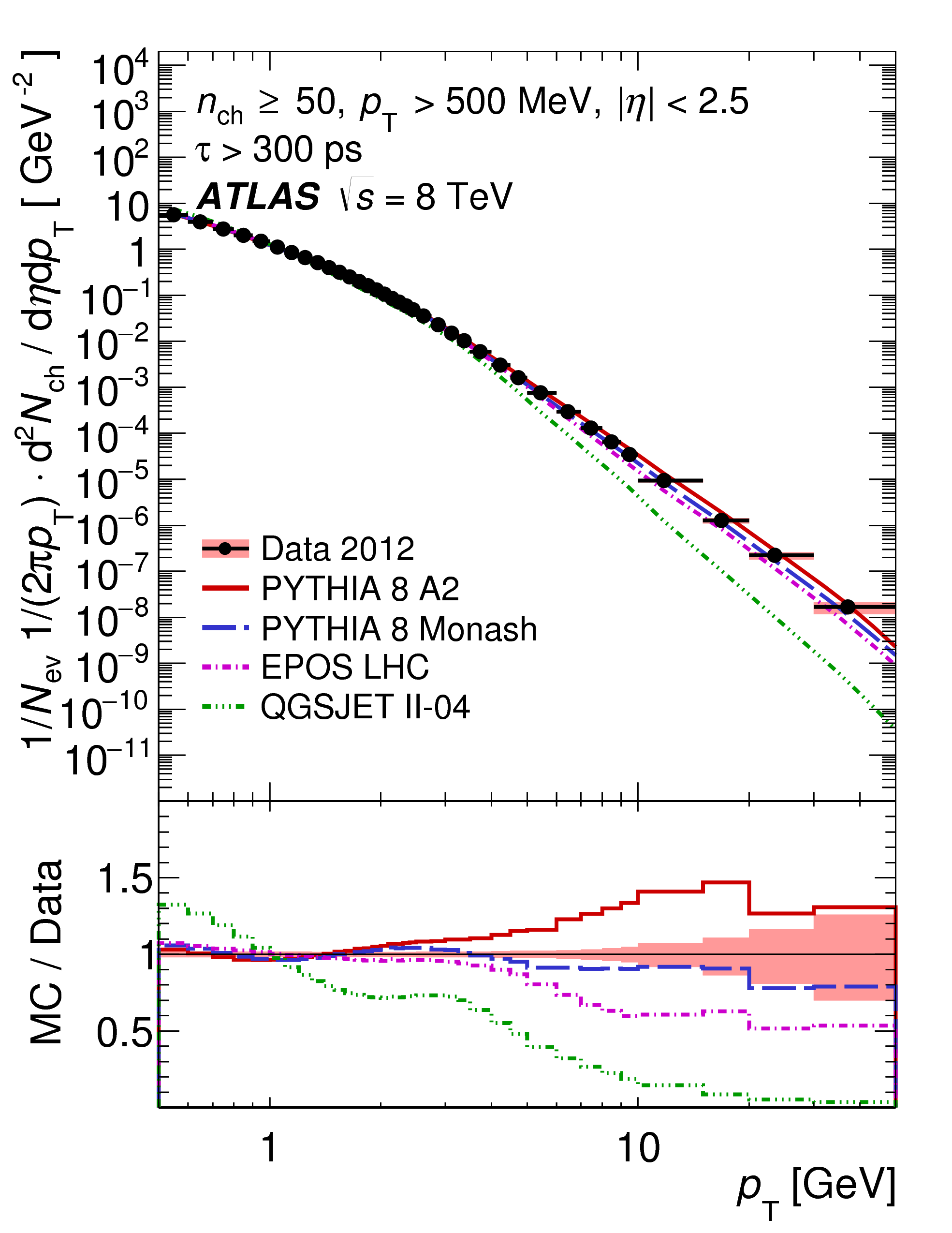}\label{fig:8sub8b}}
  \caption{ATLAS results~\cite{Aad:2016xww} on the charged-particles multiplicities as function of the leading track transverse momentum in different $\pt$ and  $n_{\rm ch}$ regions in 8~TeV $pp$ collisions: \subref{fig:8sub4b} $n_{\rm ch} \geq 2$ with $\pt{} > 100\,$MeV; and \subref{fig:8sub6b} $n_{\rm ch} \geq 6$, \subref{fig:8sub7b} $n_{\rm ch} \geq 20$ and \subref{fig:8sub8b} $n_{\rm ch} \geq 50$ with $\pt{} > 500\,$MeV.}
\label{res:8TeV:mult}
\end{figure*}

In addition, an analysis with the $\pt{} > 100\,$MeV track requirement with $n_{\rm ch} \geq 2$ was also performed by ATLAS, showing that all generators differ at forward rapidities, though within systematic uncertainties. In detail, \Pythia{}8 A2 tune and the \QGSJET{} generator describe the $\eta$ shape but fail at overall normalization. Only reasonable $\pt{}$-dependence description is obtained across models, with the \QGSJET{} generator showing largest deviations. All 8 TeV measurements are systematics-limited in precision.
Charged-particles multiplicities as function of $\pt$ in the different multiplicity phase-spaces are presented in~Figure~\ref{res:8TeV:mult}, showing a different level of agreement between models across the phase-spaces.

Event-by-event charge-particles multiplicity are best described by the \EPOS{} generator. All generators describe the $\eta$ dependence while the \QGSJET{} generator does not describe the $n_{\rm ch}$ nor $\pt$ dependence while a fair description is found by the \Pythia{}8 A2 tune.

\subsection{13 TeV Results}

\begin{figure*}[!htbp]
  \centering
    \subfigure[]{ \includegraphics[width=0.47\textwidth]{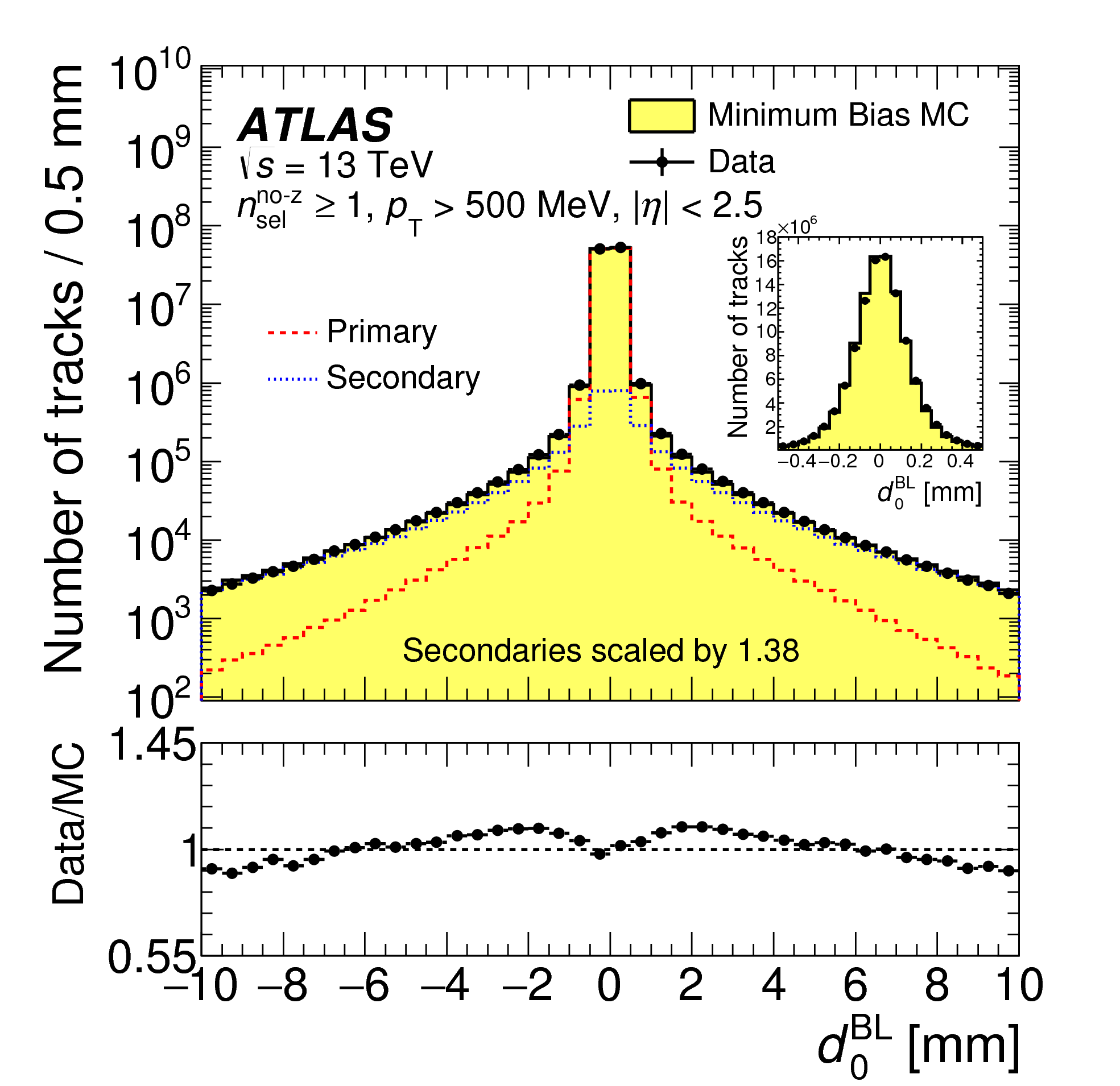}\label{fig:sub1}}
    \subfigure[]{ \includegraphics[width=0.47\textwidth]{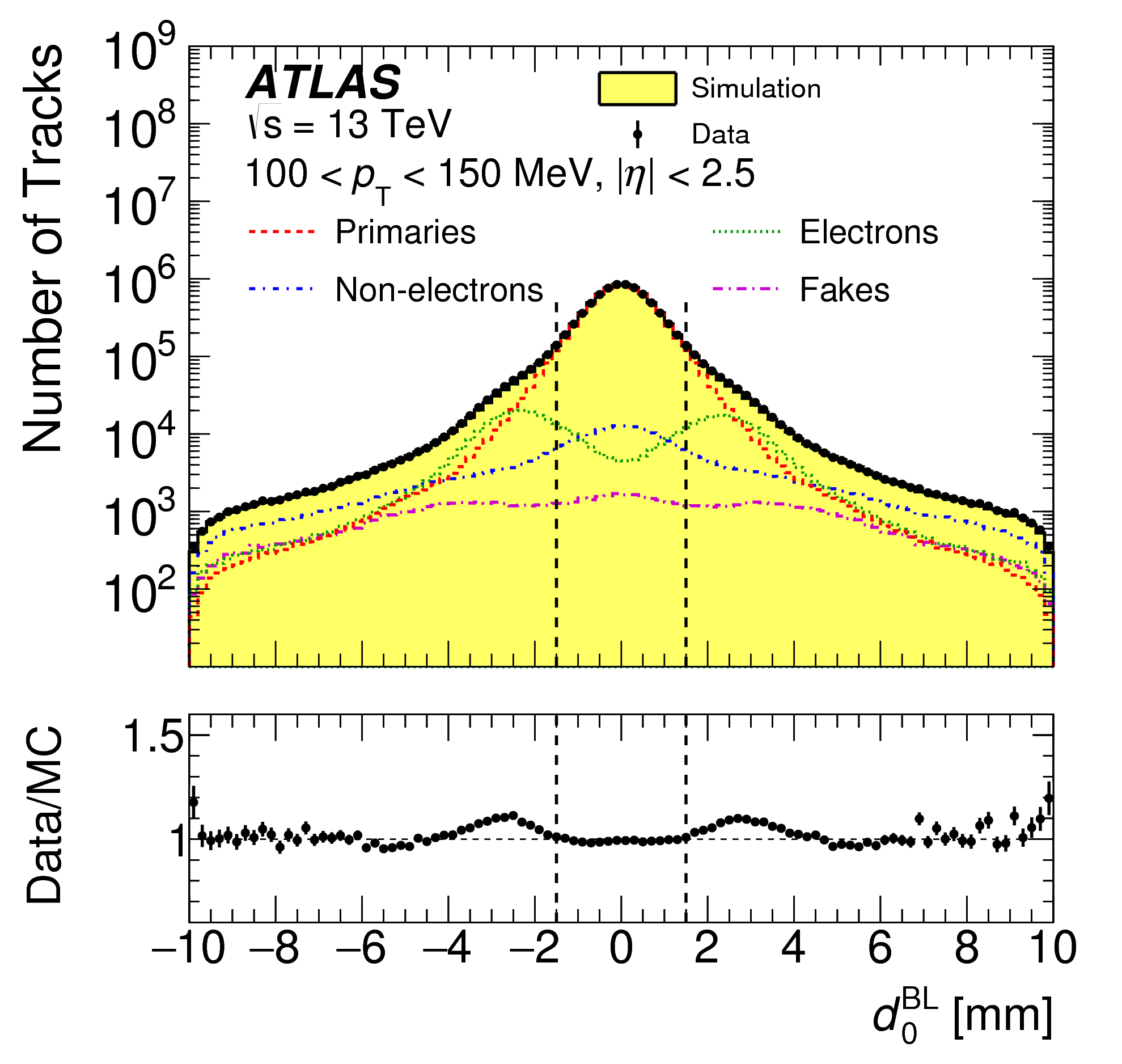}\label{fig:sub2}}
  \caption{The transverse impact parameter of tracks, defined  w.r.t. the beam line (BL) position, in the \subref{fig:sub1} $\pt{} > 500\,$MeV and \subref{fig:sub2}  $\pt{} > 100\,$MeV analysis in 13~TeV $pp$ collisions~\cite{Aad:2016xww,Aad:2016mok}.}
\label{fig:d0}
\end{figure*}

\begin{figure*}[htbp]
  \centering
  \subfigure[]{\includegraphics[width=0.47\textwidth]{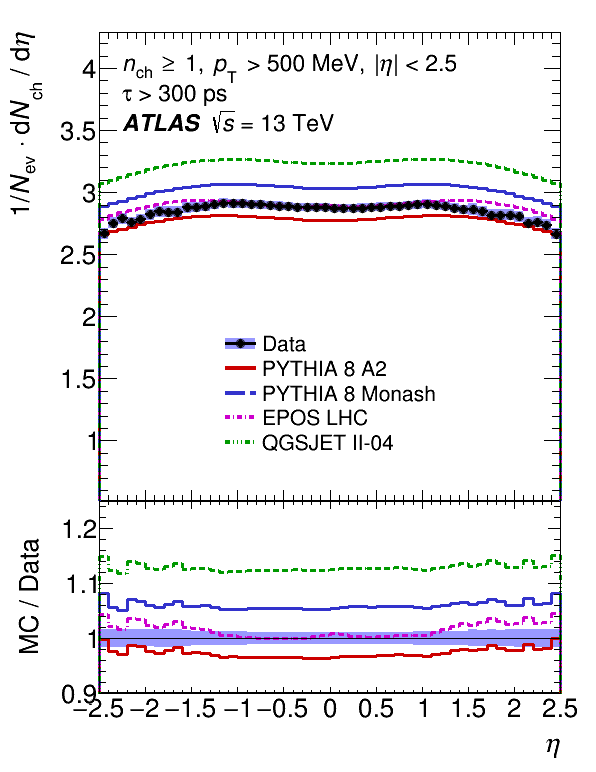}\label{fig:sub3a}}
  \subfigure[]{\includegraphics[width=0.47\textwidth]{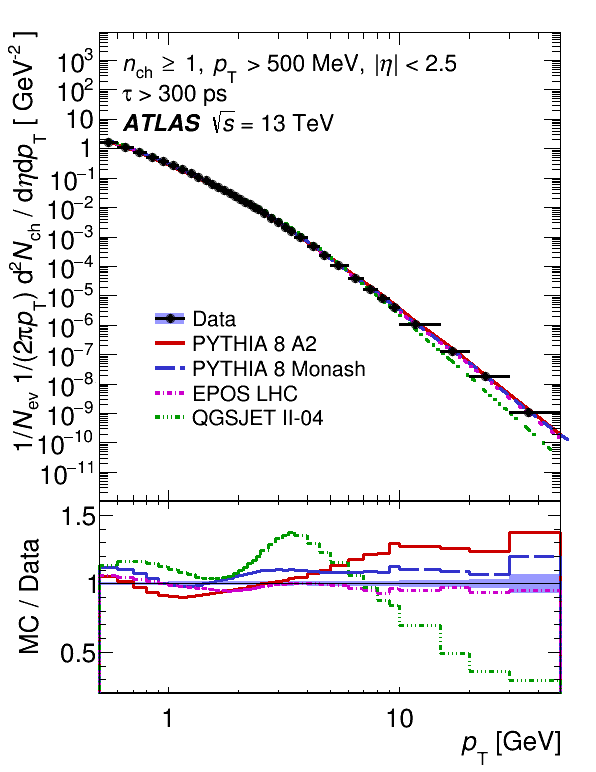}\label{fig:sub3b}}
  \caption{ATLAS results~\cite{Aad:2016mok} on the \subref{fig:sub4c} $\eta$ and \subref{fig:sub4d} $\pt{}$ spectra of charged particles in 13~TeV $pp$ collisions in the $\pt{} > 500\,$MeV and $n_{\rm ch} \geq 2$ phase space in 13~TeV $pp$ collisions.}
\label{fig:13TeV_500_pteta}
\end{figure*}

\begin{figure*}[htbp]
  \centering
  \subfigure[]{\includegraphics[width=0.47\textwidth]{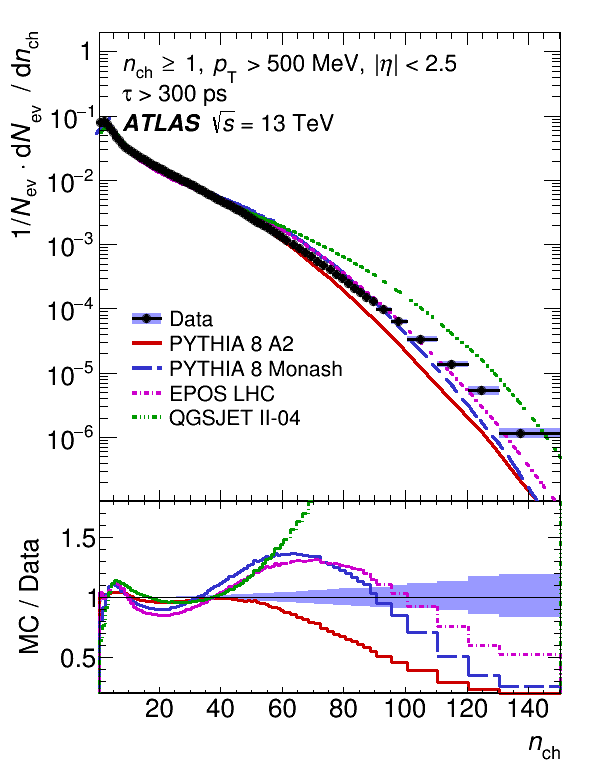}\label{fig:sub3c}}
  \subfigure[]{\includegraphics[width=0.47\textwidth]{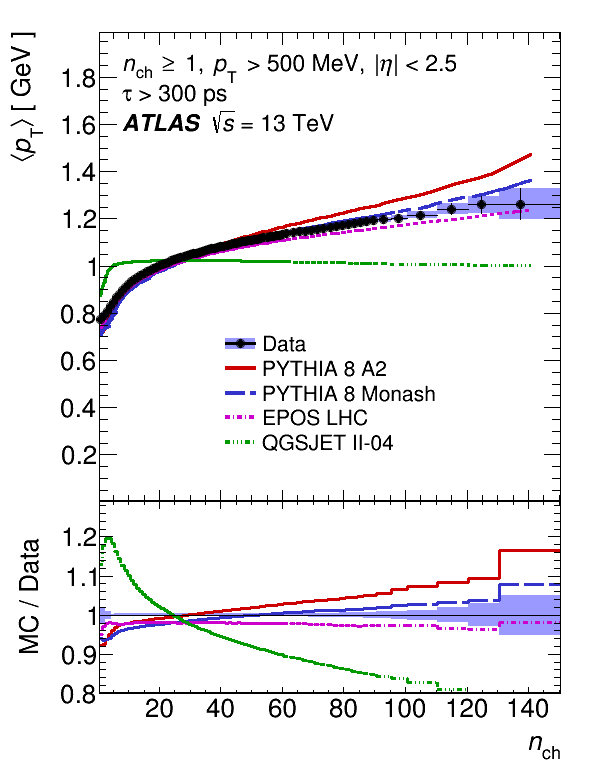}\label{fig:sub3d}}
  \caption{ATLAS results~\cite{Aad:2016mok} on the \subref{fig:sub4c} charged-particle multiplicity and the \subref{fig:sub4d} average track $\pt{}$ as function of $n_{\rm ch}$ in 13~TeV $pp$ collisions in the $\pt{} > 500\,$MeV and $n_{\rm ch} \geq 1$ phase space in 13~TeV $pp$ collisions.}
\label{fig:13TeV_500_nch}
\end{figure*}

The 13 TeV analysis for tracks with $\pt{} > 500\,$MeV in the $n_{\rm ch} \geq 1$ phase space~\cite{Aad:2016mok} is based on 9M events collected in a $170\,\mu{\rm b}^{-1}$ dataset.
The background from the beam-halo/gas interactions is negligible while the background from non-primary particles has been carefully evaluated. A~scaling factor evaluated in side-bands of the transverse impact parameter distribution was used to scale the secondary-particles fraction in simulation to match the observed yield in data.
The secondary-particles yield is then extrapolated to the analysis phase space defined as $|d_0| < 1.5$~mm 
and their contribution is subtracted from data before unfolding.
Figure~\ref{fig:d0} shows the $d_0$ distribution with contributions from photon conversions, secondary particles from hadronic interactions in material as well as fake tracks dominating the tails in the transverse impact parameter distribution.
In Figures~\ref{fig:13TeV_500_pteta} and~\ref{fig:13TeV_500_nch} the charged-particle multiplicities are displayed as function of the leading track $\pt$, $\eta$, the charged-particles multiplicity distribution is presented and the average track $\pt$ is shown a function of $n_{\rm ch}$.

\begin{figure*}[htbp]
  \centering
  \subfigure[]{\includegraphics[width=0.47\textwidth]{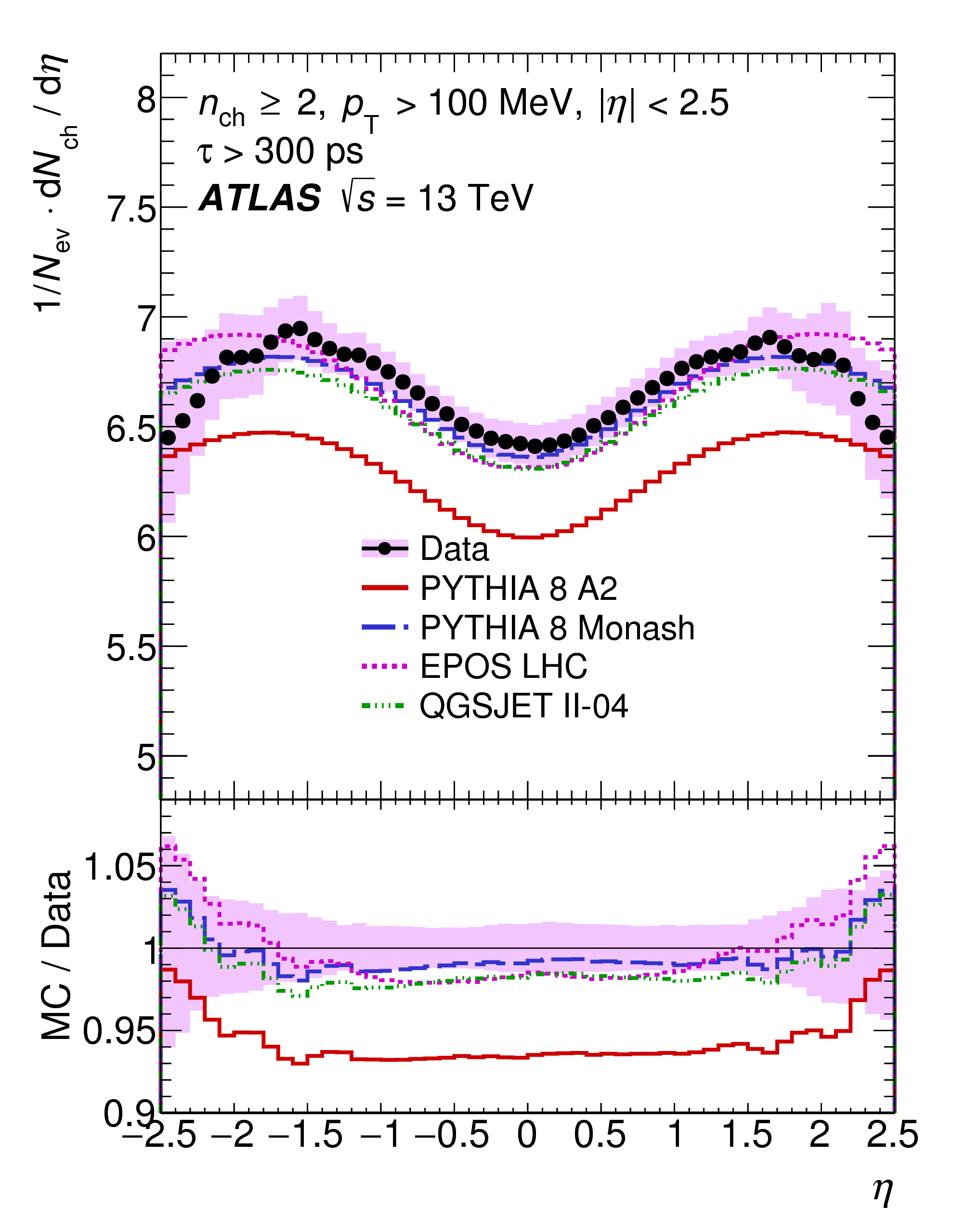}\label{fig:sub4a}}
  \subfigure[]{\includegraphics[width=0.47\textwidth]{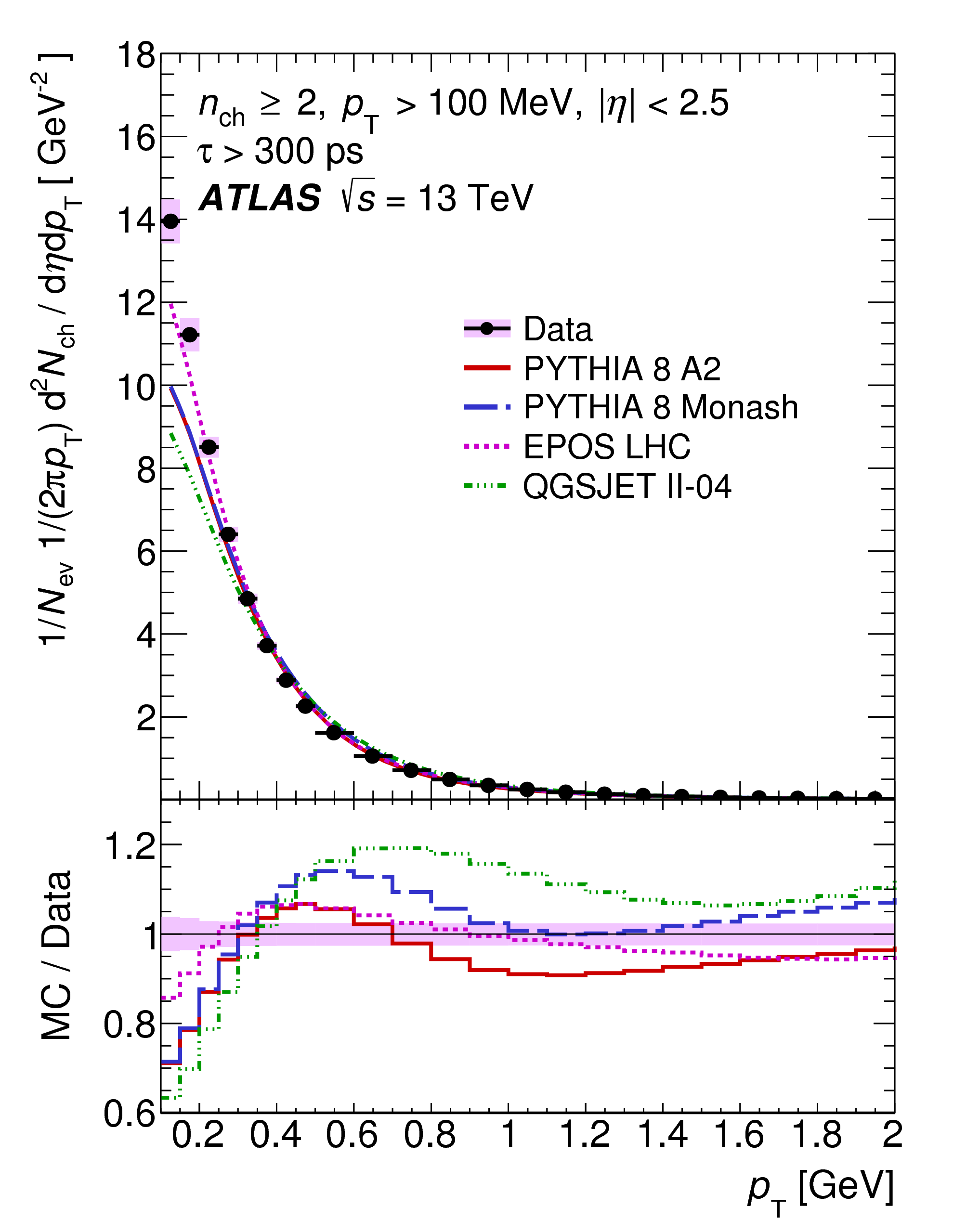}\label{fig:sub4b}}
  \caption{ATLAS results~\cite{Aaboud:2016itf} on the \subref{fig:sub4c} $\eta$ and \subref{fig:sub4d} $\pt{}$ spectra of charged-particles in~13 TeV $pp$ collisions in the $\pt{} > 100\,$MeV and $n_{\rm ch} \geq 2$ phase space in 13~TeV $pp$ collisions. }
\label{fig:13TeV_100_pteta}
\end{figure*}

\begin{figure*}[htbp]
  \centering
  \subfigure[]{\includegraphics[width=0.47\textwidth]{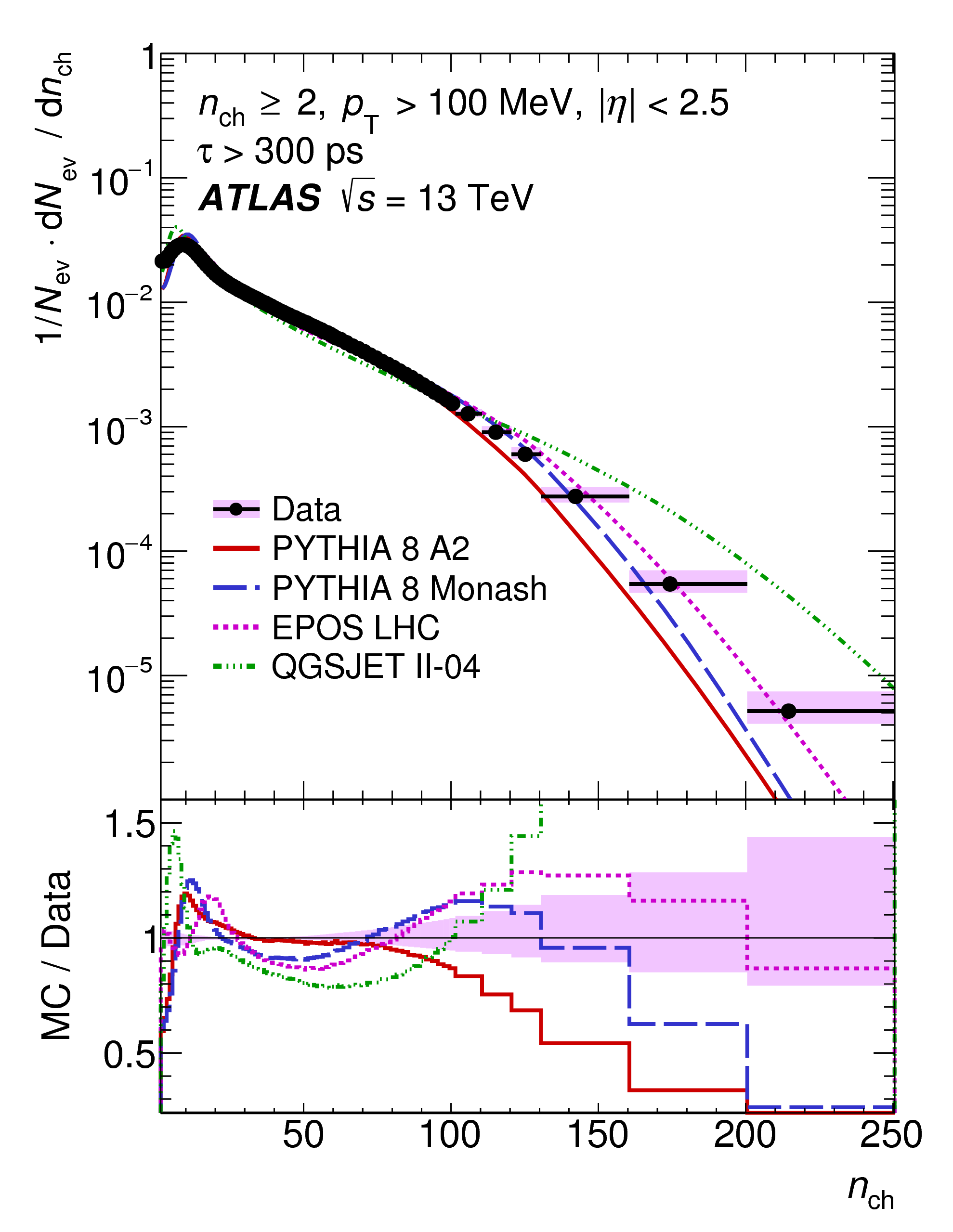}\label{fig:sub4c}}
  \subfigure[]{\includegraphics[width=0.47\textwidth]{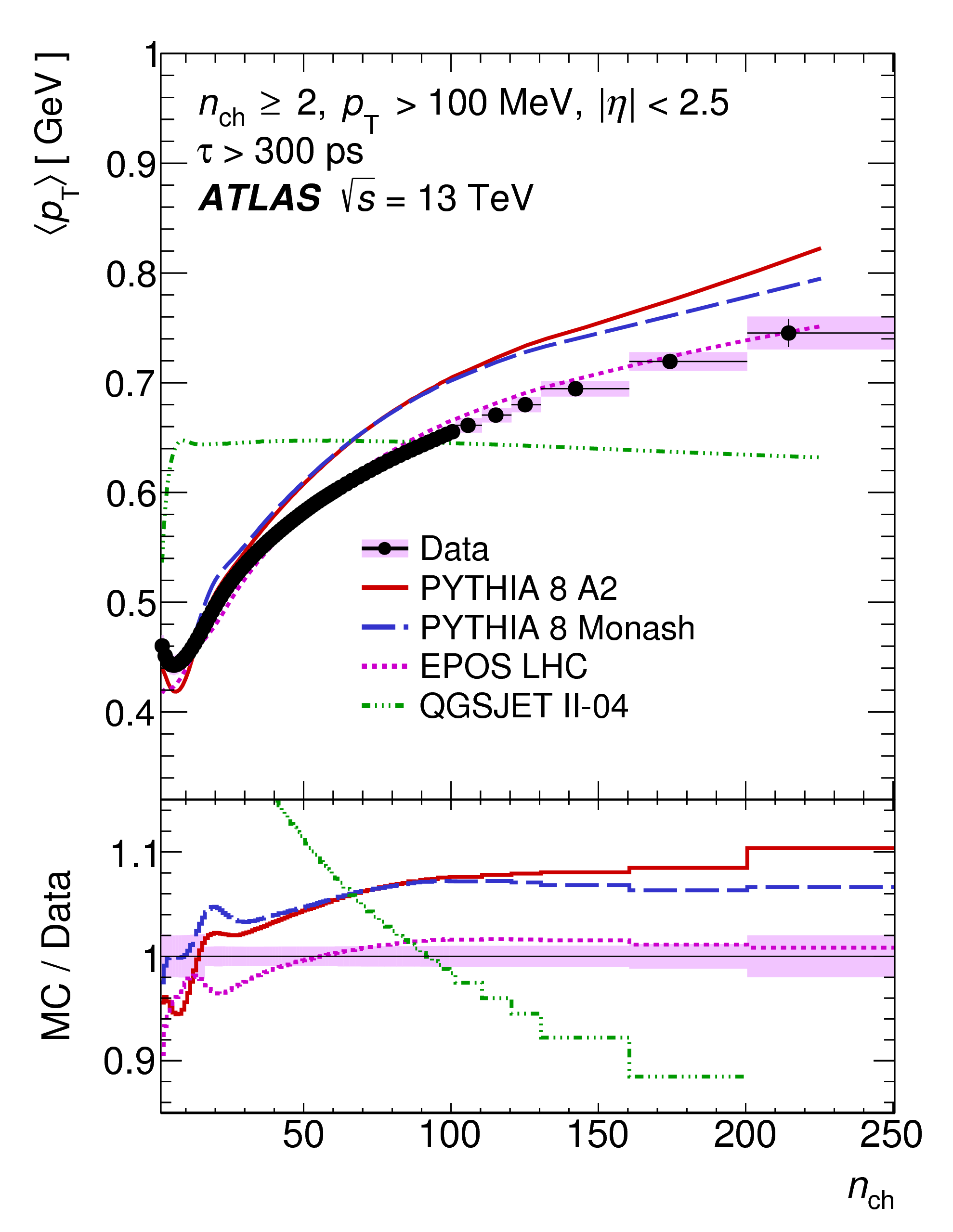}\label{fig:sub4d}}
  \caption{ATLAS results~\cite{Aaboud:2016itf} on the \subref{fig:sub4c} charged-particle multiplicity and the \subref{fig:sub4d} average track $\pt{}$ as function of $n_{\rm ch}$ in 13~TeV $pp$ collisions in the $\pt{} > 100\,$MeV and $n_{\rm ch} \geq 2$ phase space in 13~TeV $pp$ collisions.}
\label{fig:13TeV_100_nch}
\end{figure*}

ATLAS further performed a measurement of charged-particle spectra at 13 TeV for tracks with $\pt{} > 100\,$MeV in the $n_{\rm ch} \geq 2$ phase space~\cite{Aaboud:2016itf} based on 9M events collected in a $151\,\mu{\rm b}^{-1}$ dataset.
In Figures~\ref{fig:13TeV_100_pteta} and~\ref{fig:13TeV_100_nch} the charged-particle multiplicities are displayed as function of track $\pt$, $\eta$, and the multiplicity distribution and the average track $\pt$ are shown as a~function of $n_{\rm ch}$. All generators differ at forward rapidities, though within systematic uncertainties. \Pythia{}8 A2 describes shapes of distributions but fails at overall normalization. The fraction of the diffractive component and thus also the total cross-section are expected to be better described by a~future \Pythia{}8 A3 tune.
Only reasonable description of the $\pt$-dependence is seen, where the \QGSJET{} generators shows largest deviations.
\Pythia{}8 tunes and the \EPOS{} generator perform well for the multiplicity shape. The \QGSJET{} generator fails for $\langle \pt \rangle$ while the \EPOS{} generator is the best model. All generators fail at low and high $n_{\rm ch}$. ATLAS presents 13 TeV results also available in the $|\eta|<0.8$ phase-space to compare to ALICE and CMS results.

\section{Conclusions}

The ATLAS experiment presents a plethora of results on the multiplicity and spectra of charged particles produced in~8 and 13~TeV $pp$ collisions at the LHC. The new measurements provide important and tighter constraints on existing models of soft QCD interactions.
New 8~and 13~TeV results are presented, with the additional re-inclusion of strange baryons for comparison to previous measurements. Different levels of agreement to various event generators and their tunes are observed with a fair description of the 13~TeV data by the \EPOS{} generator, and \Pythia{}8 Monash and A2 tunes. However, the data hint of spectra disagreement in forward $\eta$ while all models fail at small and large multiplicities, the latter partly due to the limited level of precision of the LO generators. In addition, ATLAS also measured charged particles multiplicities per event per unit $\eta$ in central region at 900~GeV, and 2.36 and 7~TeV~\cite{Aad:2010ac}.

The author acknowledges the financial support from the project IGA no. PrF/2016/002 of Palacky University in Olomouc, Czech Republic.

\bibliography{main}{}
\bibliographystyle{plain}

\end{document}